\documentclass[prl,aps,superscriptaddress,floatfix,tightenlines,showpacs,twocolumn]{revtex4-1}
\usepackage{graphicx}
\usepackage{dcolumn}   
\usepackage{bm}        
\usepackage{amssymb}   
\usepackage{amsmath}
\usepackage{url}
\usepackage{layout}
\usepackage{color}
\usepackage{epsfig}
\usepackage{graphicx}
\usepackage{mathrsfs}\usepackage{bm}\usepackage{appendix}\usepackage{color}

\usepackage[thinlines]{easytable}
\usepackage{makecell}
\usepackage{tikz,pgf}
\usepackage{booktabs}
\usepackage{float}
\usepackage{hyperref}
\hypersetup{colorlinks,allcolors=blue}

\begin{document}

\title{The $s^\pm$-Wave Superconductivity in Pressurized La$_4$Ni$_3$O$_{10}$}

\author{Ming Zhang}
\thanks{These three authors contributed equally to this work.}
\affiliation{Department of Physics, Zhejiang Sci-Tech University, Hangzhou 310018, Zhejiang, China}

\author{Hongyi Sun}
\thanks{These three authors contributed equally to this work.}
\affiliation{Department of Physics, Southern University of Science and Technology, Shenzhen 518055, China}
\affiliation{Shenzhen Key Laboratory of Advanced Quantum Functional Materials and Devices, Southern University of Science and Technology, Shenzhen 518055, China}

\author{Yu-Bo Liu}
\thanks{These three authors contributed equally to this work.}
\affiliation{School of Physics, Beijing Institute of Technology, Beijing 100081, China}

\author{Qihang Liu}
\affiliation{Department of Physics, Southern University of Science and Technology, Shenzhen 518055, China}
\affiliation{Shenzhen Key Laboratory of Advanced Quantum Functional Materials and Devices, Southern University of Science and Technology, Shenzhen 518055, China}

\author{Wei-Qiang Chen}
\email{chenwq@sustech.edu.cn}
\affiliation{Department of Physics, Southern University of Science and Technology, Shenzhen 518055, China}
\affiliation{Shenzhen Key Laboratory of Advanced Quantum Functional Materials and Devices, Southern University of Science and Technology, Shenzhen 518055, China}

\author{Fan Yang}
\email{yangfan\_blg@bit.edu.cn}
\affiliation{School of Physics, Beijing Institute of Technology, Beijing 100081, China}

\begin{abstract}
Recently, evidence of superconductivity (SC) has been reported in pressurized La$_4$Ni$_3$O$_{10}$. Here we study the possible pairing mechanism and pairing symmetry in this material. Through fitting the density-functional-theory band structure, we provide a six-orbital tight-binding model. In comparison with the band structure of La$_3$Ni$_2$O$_7$, the additional non-bonding $d_{z^2}$ band is importance to the pairing mechanism here. When the multi-orbital Hubbard interactions are included, our random-phase-approximation based study yields an $s^{\pm}$-wave pairing. The dominant FS nesting with nesting vector $\mathbf{Q}_1\approx (\pi,\pi)$ is between the $\gamma$-pocket contributed by the bonding $d_{z^2}$ band top and the $\alpha_1$-pocket contributed by the non-bonding $d_{z^2}$ band bottom, leading to the strongest pairing gap amplitude and opposite gap signs within the two regimes. The dominant real-space pairing is the interlayer pairing between the $d_{z^2}$ orbitals. We have also studied the doping dependence of the pairing symmetry and $T_c$.
\end{abstract}\maketitle


{\bf Introduction:}
The recent discovery of superconductivity (SC) in the nickel-based family~\cite{Li2019,Li2020,Zeng2020,Osada2020,Osada2021,Pan2021,Zeng2022,WangM3}, especially the high-temperature SC near 80 K in the bilayer nickelate La$_3$Ni$_2$O$_7$ under pressures~\cite{Sun2023,Cheng2023,Wen2023,Yuan2023,Zhou2023,Qi2023,Cheng20232,Cheng20233,Liu2023,LSun2023,Guo2023,JFMitchell,ZXShen,MHepting,WXie,DLF,MWang,XWW,ZChen,XHChen}, has raised a surge of interest to explore the electron correlation and pairing nature in this family~\cite{Yang,Lu2023,Dagotto2,Wang20232,Eremin2023,Kuroki20232,Hu2023,Zhang2023,Werner2023,Leonov2023,ZhangYH2023,Si2023,Su2023,ZhangFC,ZhangFC2,WKu,Kuroki3,YaoDX2,WengZY,WuC2,Hirschfeld2023,ZhangYH2,Wahl,Bohrdt,Botana2,Kotliar,Kuroki4,Wehling,Dagotto3,Bohrdt2,LiW,ChenH,LuZY2,HuangB,W´uW,Eremin,JHu,Hirschfeld,Chistyakov,TXiang,Savrasov,ZXLi,Grusdt2023}. La$_3$Ni$_2$O$_7$ belongs to the Ruddlesden-Popper (RP) phase with formula La$_{n+1}$Ni$_n$O$_{3n+1}$~\cite{Beznosikov2000,Lacorre1992}, which consists of n layers of perovskite-type LaNiO$_3$, separated by a single rocksalt-type LaO layer along the c-axis direction. While $n=2$ is for La$_3$Ni$_2$O$_7$, the $n=3$ member of this family is $R_4$Ni$_3$O$_{10}$ ($R$=La, Pr, Nd). Recently, clear drops in resistance of La$_4$Ni$_3$O$_{10}$ under pressure at about 20-30 K were observed~\cite{Zhangar,Sakakibaraar,Liar,Zhuar,Mukuda,Botana}, indicating signatures of SC.

La$_4$Ni$_3$O$_{10}$ hosts a quasi-2D crystal structure, with approximate unit cell comprising three NiO$_2$ layers interconnected by the Ni-O-Ni $\sigma$
bond. Under high pressure, there's a structural transition from monoclinic $P2_1/a$ to tetragonal $I4/mmm$. It takes the 164.8° Ni-O-Ni angle between adjacent octahedra layers forced to 180.0° in the high pressures phase, reminiscent of the bilayer La$_3$Ni$_2$O$_{7}$~\cite{Sun2023,Cheng2023,Wen2023,Yuan2023,Zhou2023,Qi2023,Cheng20232,Cheng20233,Liu2023,LSun2023,Guo2023}. While the NiO$_2$ plane in La$_4$Ni$_3$O$_{10}$ is isostructural with the CuO$_2$ plane in the cuprates, the nominal valence state of Ni is +2.67, leading to the electron configuration $d^{7.33}$, which is different from the $d^9$ state in the cuprates. This electron configuration is also different from that in the infinite-layer nickelates and in the Ni$^{2.5+}$ ($d^{7.5}$) state found in bilayer La$_3$Ni$_2$O$_7$. Particularly, the filling fractions of both $3d$ orbitals in La$_4$Ni$_3$O$_{10}$ are near $1/3$, which is quite different from that in La$_3$Ni$_2$O$_{7}$~\cite{Sun2023}.

The density-functional-theory (DFT) based calculations suggest that the low-energy degrees of freedom in La$_4$Ni$_3$O$_{10}$ are mainly the Ni- $3d_{z^2}$ and $3d_{x^2-y^2}$ orbitals~\cite{Sakakibaraar,Dessau2017,Wang20231}. Similar with La$_3$Ni$_2$O$_{7}$, the interlayer coupling in La$_4$Ni$_3$O$_{10}$ are mainly realized through the strong hybridization between the Ni-$3d_{z^2}$ orbital and the O-$p_z$ orbital on the inner apical oxygen atoms~\cite{YangF2023,WuC2023,Wang20231}. Such strong interlayer coupling renders that the $3d_{z^2}$- orbital dominant bands are split into the bonding-, non-bonding- and anti-bonding- bands~\cite{Sakakibaraar}. Similarly with La$_3$Ni$_2$O$_7$, the pressure lifts up the top of the bonding 3$d_{z^2}$ band to cross the Fermi level, forming into a hole pocket near the Brillouin zone (BZ) corner M$(\pi,\pi)$ point~\cite{Sakakibaraar,Wang20231,QHW,DXY,Leonov,HLi,ZLu}, which might be important for the emergence of SC in the pressurized system~\cite{WuC4}. The anti-bonding 3$d_{z^2}$ component is well above the Fermi level.  The non-bonding 3$d_{z^2}$ band, which significantly hybridizes with the 3$d_{x^2-y^2}$ bands, is a new band absent in La$_3$Ni$_2$O$_7$~\cite{Sakakibaraar,Wang20231}. As this band has a local bottom at the $\Gamma$-point and the local bottom is very near the Fermi level, band structures at the $\Gamma$-point may host an electron pocket~\cite{Dagotto,Leonov,ZLu,HLi} or not~\cite{QHW,DXY} from different research groups,depending on the band details. This band might also take an important role in the pairing mechanism of pressurized La$_4$Ni$_3$O$_{10}$~\cite{Dagotto}. 

In this paper, we study the pairing mechanism and pairing symmetry in pressurized La$_4$Ni$_3$O$_{10}$ through standard random-phase-approximation (RPA) approach. We start from a six-orbital tight-binding (TB) model which well fits our DFT band structure. After adding multi-orbital Hubbard interactions into the model, our RPA result provides an $s^{\pm}$-wave pairing. The dominant FS nesting is between the $\gamma$-pocket centering around the $M$-point contributed by the bonding $d_{z^2}$ band and the $\alpha_1$-pocket centering around the $\Gamma$-point contributed by the non-bonding $d_{z^2}$ band. The pairing gap is mainly distributed on the two pockets, with opposite gap signs on the two pockets. The real-space pairing pattern is dominated by interlayer 3$d_{z^2}$ pairing. While the pairing symmetry maintains $s^\pm$ for low doping levels, the $T_c$ arrives at its maximum at slight electron doping. 

\begin{figure}[htbp]
\centering
\includegraphics[width=0.5\textwidth]{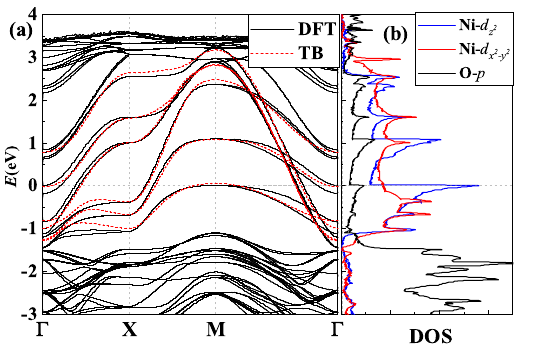}
\caption{(color online) (a) The DFT(black solid line) and TB(red dashed line) band structure of $\text{La}_4\text{Ni}_3\text{O}_{10}$ under the pressure of 40Gp. The experimental refined lattice constants are adopted. (b) The DOS for different orbital components of the DFT band of $\text{La}_4\text{Ni}_3\text{O}_{10}$ under 40Gp. The bule, red and black lines represent Ni-$d_{z^2}$,Ni$-d_{x^2-y^2}$,O$-p$ orbitals, respectively.}
\label{dft}
\end{figure}

{\bf Band Structure and TB model:} To study the band structure of pressurized La$_4$Ni$_3$O$_{10}$, we adopt the tetragonal $I4/mmm$ conventional cell with six Ni-atoms. Our DFT calculations utilized the projector-augmented wave (PAW) pseudo-potentials with the exchange–correlation of the Perdew–Burke–Ernzerhof and the GGA approach, as implemented in the Vienna ab-initio Simulation Package (VASP)\cite{dft1,dft2,dft3}. To account for the correlation effects of 3d electrons in Ni atoms, we employed GGA+U scheme\cite{dft4} and setting U to 3.5 eV [arXiv:2309.01148(2023)]. The cutoff energy for the plane-wave basis was set to 600 eV. The reciprocal space was sampled using a $20\times20\times3 k$ mesh for structural optimization and self-consistent calculations. The lattice constants measured experimentally at 40GPa were utilized [arXiv:2311.16763], and the atomic positions were subsequently relaxed until the atomic force on each atom was less than $10^{-3}$ eV/$\dot{\text{A}}$. 

The obtained band structure is shown in Fig.~\ref{dft}(a), and the density of state (DOS) that contributed by different orbital components, is shown in Fig.~\ref{dft}(b). Fig.~\ref{dft}(b) shows that the low-energy DOS is mainly contributed by the two Ni-$3d$ orbitals, with the $3d_{z^2}$ dominating the $3d_{x^2-y^2}$ in the contribution. To acquire a TB description of the DFT band structure, we constructed maximally localized Wannier representations\cite{dft5} by projecting the Bloch states (with an 20×20×3 $k$ mesh) from the DFT calculations onto the $3d_{z^2}$ and $3d_{x^2-y^2}$ orbitals. As depicted in the Supplementary Materials (SM), the band structure obtained from the complete Wannier representations agrees very well with that from DFT calculation across the entire energy range of interest.

\begin{figure}[htbp]
\centering
\includegraphics[width=0.5\textwidth]{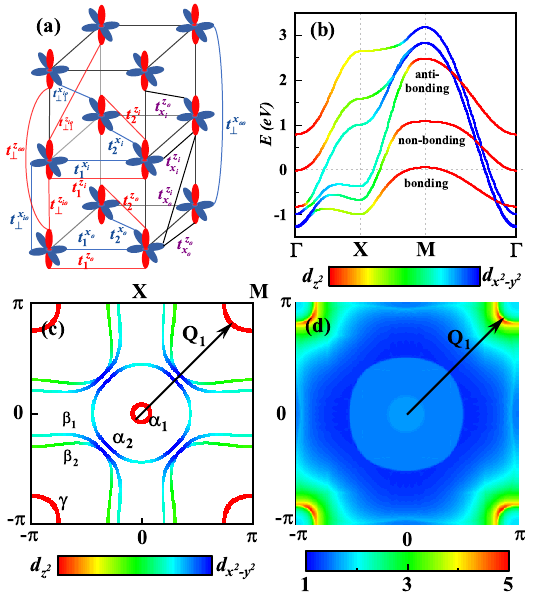}
\caption{(color online) The trilayer six-orbital TB model of La$_4$Ni$_3$O$_{10}$ under 40Gp and spin susceptibility. (a) Schematic of trilayer La$_4$Ni$_3$O$_{10}$ lattice with hopping parameters. The red(bule) pattern represents Ni$-d_{z^2}(d_{x^2-y^2})$ orbital. The values of hopping parameters are listed in Tab.I. (b) The band structure along the high symmetry lines. (c) FS in the first Brillouin zone. The five pockets are labeled by $\alpha_1,\alpha_2,\beta_1,\beta_2$ and $\gamma$, respectively. The FS-nesting vector is marked by $\mathbf{Q_1}$. The colorbar in (b-c) indicates the orbital weight of $d_{x^2-y^2}$ and $d_{z^2}$. (d) The distribution of the RPA-renormalized spin susceptibility $\chi^{(s)}(\mathbf{q})$ for $U=0.9$ eV. The distribution maximum of the spin susceptibility is just $\mathbf{Q_1}$.}
\label{tb}
\end{figure}

To facilitate subsequent studies involving electron interactions, we neglect the coupling between the upper three layers and the lower three layers within a unit cell, and by taking each three layers as a unit cell we extract from the Wannier representations the following six-orbital TB model up to the third-nearest
neighbor hopping,
\begin{eqnarray}
H_{\text{TB}}=\sum_{i\mu\sigma}\varepsilon_{i\mu}c^{\dagger}_{i\mu\sigma}c_{i\mu\sigma}+\sum_{ij,\mu\nu,\sigma}t_{ij,\mu\nu}c^{\dagger}_{i\mu\sigma}c_{j\nu\sigma}+h.c.
\end{eqnarray}
Here $i/j$ denote site indices, i.e. the combined in-plane coordinate and layer indices. $\mu/\nu$ label orbital, and $\sigma$ labels spin. $\varepsilon_{i\mu}$ represents the on-site energy of orbital $\mu$ at the site $i$. The relevant hopping integrals $t_{ij,\mu\nu}$ up to the NNN bonds are illustrated in Fig.~\ref{tb}(a), with their values listed in Tab.~I. Here the notation $x/z$ indicate the $d_{x^2-y^2}/d_{z^2}$ orbitals, $i/o$ represent the inner/outer layers, $1/2/3$ indicate NN/NNN/TNN intralayer hoppings and $_\perp/_{\perp1}$ mean interlayer hoppings. Similar with La$_3$Ni$_2$O$_{7}$, for the $d_{z^2}$ orbital, the interlayer coupling is strong but the intralayer coupling is weak; while for the $d_{x^2-y^2}$ orbital, the interlayer couplings is weak but the intralayer coupling is strong\cite{Yao20232,Wang20233,SuG2,Kuroki5}. 

\begin{table}[!h]
\label{tab:1}
\centering
\caption{TB parameters of the trilayer La$_4$Ni$_3$O$_{10}$. In the superscript and subscript, $x(z)$ represents the $d_{x^2-y^2}(d_{z^2})$ orbit, $o(i)$ represents the outer(inner) layer, $\perp$ represents interlayer hopping, and $1,2,3$ represent the nearest, second nearest neighbors and third-nearest
neighbor, respectively. The hoppings $t$ are demonstrated in Fig.~\ref{tb}(a). The unit of all parameters is $eV$.}
\begin{tabular}{ccccccc}
  \hline\hline
  $t_{1}^{x_o}$ & $t_{2}^{x_o}$ & $t_{1}^{z_o}$ & $t_{2}^{z_o}$ & $t_{x_i}^{z_o}$ & $t_{x_o}^{z_o}$ & $\epsilon_{x_o}$ \\
  \hline
  -0.512 & 0.067 & -0.149 & -0.014 & 0.037 & -0.277 & 0.184 \\
  \hline
  $t_{1}^{x_i}$ & $t_{2}^{x_i}$ & $t_{1}^{z_i}$ & $t_{2}^{z_i}$ & $t_{x_i}^{z_i}$ & $t_{x_o}^{z_i}$ & $\epsilon_{x_i}$ \\
  \hline
  -0.521 & 0.069 & -0.160 & -0.021 & -0.292 & 0.035 & 0.480  \\
  \hline
  $t_{\perp}^{x_{io}}$& $t_{\perp1}^{x_{io}}$  & $t_{\perp}^{z_{io}}$ & $t_{\perp1}^{z_{io}}$ & $t_{\perp}^{x_{oo}}$ & $t_{\perp}^{z_{oo}}$ & $\epsilon_{z_i}$ \\
  \hline
  0.010 & 0.002 & -0.705 & 0.038 & -0.003 & -0.099 & 0.312 \\
     \hline  
  $t_{3}^{x_{o}}$& $t_{3}^{x_{i}}$  & $t_{3}^{z_{o}}$ & $t_{3}^{z_{i}}$ & $t_{\perp1}^{x_{oo}}$ & $t_{\perp1}^{z_{oo}}$ & $\epsilon_{z_o}$ \\
  \hline
  -0.062 & -0.061 & -0.019 & 0.021 & -0.001 & -0.012 & 0 \\
     \hline\hline
\end{tabular}
\end{table}

The obtained band structure for this TB model is shown in Fig.~\ref{tb}(b), with the associate FS shown in Fig.~\ref{tb}(c). The orbital component is marked by the color. The strong interlayer hoppings in combination with the weak intralayer hoppings for the $d_{z^2}$ orbital renders that the $d_{z^2}$ bands are split into the bonding, non-bonding and anti-bonding bands. These $d_{z^2}$-dominant bands are mixed with $d_{x^2-y^2}$ component through hybridization. The top of the bonding $d_{z^2}$ band crosses the Fermi level, forming  the hole-like $\gamma$-pocket centred around the BZ corner M-point, as shown in Fig.~\ref{tb}(c). The anti-bonding $d_{z^2}$ component by itself is completely above the Fermi level (see the red part). Besides, it hybridizes with the $d_{x^2-y^2}$- component to form a new band crossing the Fermi level, leading to the hole-like $\beta_1$-pocket around M, whose dominant orbital component is $d_{x^2-y^2}$. The non-bonding $d_{z^2}$ band is near the Fermi level and significantly hybridizes with the $d_{x^2-y^2}$-component. This band has a local bottom at the $\Gamma$-point, which slightly crosses the Fermi level, forming the electronic-like $\alpha_1$-pocket centred around the BZ centre $\Gamma$-point. Besides, this band also contributes a large hole-like $\beta_2$-pocket with comparable $d_{z^2}$ and $d_{x^2-y^2}$ components. In comparison with La$_3$Ni$_2$O$_{7}$, the non-bonding $d_{z^2}$ band is new.  In addition, an extra electron-like $\alpha_2$-pocket centering around the $\Gamma$-point is contributed by the $d_{x^2-y^2}$ orbital.

{\bf Interaction and Spin Susceptibility:} We adopt the following multi-orbital Hubbard interaction,
\begin{align}\label{model}
H_{int}&=U\sum_{i\mu}n_{i\mu\uparrow}n_{i\mu\downarrow}+
V\sum_{i,\sigma,\sigma^{\prime}}n_{i1\sigma}n_{i2\sigma^{\prime}}+J_{H}\sum_{i\sigma\sigma^{\prime}} \nonumber\\
&\Big[c^{\dagger}_{i1\sigma}c^{\dagger}_{i2\sigma^{\prime}}c_{i1\sigma^{\prime}}c_{i2\sigma}+(c^{\dagger}_{i1\uparrow}c^{\dagger}_{i1\downarrow}c_{i2\downarrow}c_{i2\uparrow}+h.c.)\Big]
\end{align}
Here, $U$, $V$, and $J_H$ denote the intra-orbital, inter-orbital Hubbard repulsion, and the Hund's rule coupling (and the pair hopping) respectively, which satisfy the relation $U=V+2J_H$. For the subsequent calculations, we fix $J_H = U/6$. We employ the multi-orbital RPA approach to treat this Hamiltonian\cite{RPA1,RPA2,RPA3,RPA4,RPA5,RPA6,RPA7,RPA8}.

We define the following bare susceptibility
 \begin{align}\label{chi0}
 \chi^{(0)pq}_{st}(\bm{k},\tau)\equiv
 &\frac{1}{N}\sum_{\bm{k}_1\bm{k}_2}\left\langle
 T_{\tau}c_{p}^{\dagger}(\bm{k}_1,\tau)
 c_{q}(\bm{k}_1+\bm{k},\tau)\right.                      \nonumber\\
 &\left.\times c_{s}^{\dagger}(\bm{k}_2+\bm{k},0)
 c_{t}(\bm{k}_2,0)\right\rangle_0.
 \end{align}
 The spin (s) and charge (c) susceptibilities renormalized in the RPA level are given by\cite{RPA1,RPA2,RPA3,RPA4,RPA5,RPA6,RPA7,RPA8}
\begin{align}\label{chisce}
 \chi^{(s,c)}(\bm{k},\tau)=[I\mp\chi^{(0)}(\bm{k},\tau)
 U^{(s,c)}]^{-1}\chi^{(0)}(\bm{k},\tau).
\end{align}
Usually, repulsive interactions suppress $\chi^{(c)}$ but enhance $\chi^{(s)}$. There exist a critical strengths $U_c^{(c/s)}$, exceeding which $\chi^{(c/s)}$ diverges, leading to the charge/spin-density wave (CDW/SDW). Usually we have $U_c^{(c)}>U_c^{(s)}\equiv U_c$. For the present band structure $U_c\approx 1.2$ eV. 

We further define the static spin susceptibility matrix $\chi^{(s)}_{\bm{k}}(pq, st)\equiv\chi^{(s)pq}_{st}(\bm{k},i\omega=0)$, where $\omega$ is the Matsubara frequency. The largest eigenvalue of this Hermitian matrix for each momentum $\bm{k}$ is defined as $\chi^{(s)}(\bm{k})$, which represents the eigen spin susceptibility in the strongest channel. The distribution of $\chi^{(s)}(\bm{k})$ over the BZ is shown in Fig.~\ref{tb}(d) for $U=0.9$ eV$<U_c$. Notably, the strongest susceptibility locates near the momentum $\mathbf{Q}_1\approx(\pi,\pi)$, which is the nesting vector between the $\alpha_1$-pocket and $\gamma$-pocket, as shown in Fig.~\ref{tb}(c).

{\bf Pairing Symmetry and $T_c$:} When $U<U_c$, through exchanging spin/charge fluctuations represented by $\chi^{(s/c)}$, an effective interaction $V^{\alpha\beta}(\bm k,\bm q)$ \cite{RPA1,RPA2,RPA3,RPA4,RPA5,RPA6,RPA7,RPA8} can be obtained, which can mediate the pairing instability. Then via the mean-field treatment with the obtained effective Hamiltonian, one can acquire a self-consistent pairing gap equation, which after linearized near $T_c$ provides the linearized gap equation. Choosing a thin energy shell $\pm\Delta E$ near the Fermi level and integrating the high energy part, the linearized gap equation after discretized on the lattice takes the form of an eigenvalue problem of the effective interaction matrix.  The $T_c$ is related to the largest eigenvalue $\lambda$ of this matrix via $T_c\propto e^{-1/\lambda}$, while the pairing symmetry is determined by the eigen vector corresponding to this eigenvalue.

\begin{figure}[htbp]
\centering
\includegraphics[width=0.5\textwidth]{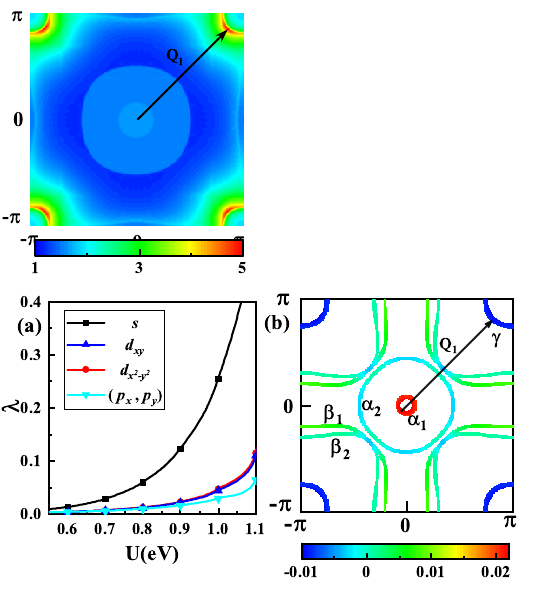}
\caption{(color online) (a) The largest pairing eigenvalue $\lambda$ of the various pairing symmetries as function of the interaction strength $U$ with fixed $J_H = U/6$. (b) The distribution of the leading $s$-wave pairing gap functions on the FS for $U=0.9$ eV. The two pockets connected by the nesting vector $\mathbf{Q_1}$ host opposite gap signs.}
\label{sc}
\end{figure}

The $U$ dependences of the largest pairing eigenvalue $\lambda$ for different pairing symmetries, including the $s$-wave, $d_{xy}$-wave, $d_{x^2-y^2}$-wave, and degenerate $(p_x,p_y)$-wave, are shown in Fig. \ref{sc}(a). Consequently, the $s$-wave is the leading pairing symmetry and dominates other ones, similar with the case in pressurized La$_3$Ni$_2$O$_{7}$\cite{LuZY,YouYZ,WuC3,Dagotto2023,Dagottoprb,ZhouT,YangYF}. The calculated pairing strength $\lambda$ ($\sim$ 0.12) of the $s^{\pm}$-wave gap structure is smaller than that in pressurized La$_3$Ni$_2$O$_{7}$ ($\sim$ 0.3) for the same $U=0.9$ eV ~\cite{Dagotto3}, which is consistent with experiments.

The distribution of the relative gap function is shown on the FS in Fig.~\ref{sc}(b). Consequently, the $\alpha_1$- and $\gamma$- pockets are distributed with the strongest pairing amplitude, with their gap signs opposite. The two pockets are just connected by the strongest nesting vector $\mathbf{Q}_1$ shown in Fig.~\ref{tb}(c), which is also reflected in the strongest spin susceptibility peak shown in Fig.~\ref{tb} (d). This $s$-wave pairing is just the $s^{\pm}$-wave pairing, similar with that in the Fe-based SC. When this $\bm {k}$-space pairing gap function is Fourier transformed to the real space, we obtain a dominant interlayer pairing of the $d_{z^2}$ orbital, as shown in the Supplementary Materials (SM). This is consistent with the fact that the dominant orbital component of the two pockets distributed with the strongest pairing amplitude, i.e. the $\alpha_1$- and $\gamma$- pockets, is $d_{z^2}$.  

\begin{figure}[H]
\centering
\includegraphics[width=0.5\textwidth]{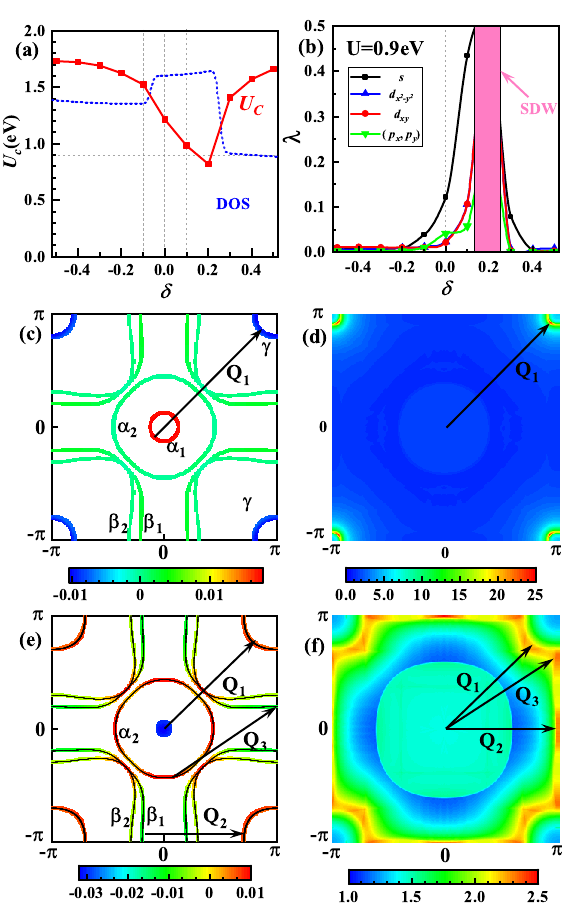}
\caption{(color online) Doping-dependent study for La$_4$Ni$_3$O$_{10}$. (a) The critical interaction strength $U_c$ (red solid line) as function of the doping $\delta$. The DOS (blue dashed line) is also plotted for comparison.  (b) The largest pairing eigenvalue $\lambda$ of the various pairing symmetries as function of the doping $\delta$ for $U=0.9$ eV. The pink area indicates SDW. (c-d) The distributions of the leading $s^{\pm}$-wave pairing gap functions (c) on the FS and spin susceptibility (d) in the BZ for electron doping $\delta=0.1$ and $U=0.9$ eV. In (c), $\mathbf{Q_1}$ marks the nesting vector. In (d), the distribution peak of the spin susceptibility is just located at $\mathbf{Q_1}$. (e-f) The distribution of the leading $s$-wave pairing gap function within an energy shell of $\pm 0.01$ eV near the ermi level (e) and spin susceptibility (f) in the BZ for hole doping $\delta=-0.1$ and $U=0.9$ eV. In (e), $\mathbf{Q_1}$ marks the ``virtual'' nesting vector, and $\mathbf{Q_2}$ and $\mathbf{Q_3}$ are true nesting vector. In (f), the distribution peaks of the spin susceptibility are just located at $\mathbf{Q_1}$, $\mathbf{Q_2}$ and $\mathbf{Q_3}$, respectively. }
\label{dop}
\end{figure}

The doping $\delta$-dependence of $U_c$ and DOS is shown in Fig.~\ref{dop}(a). The maximal DOS locates at $\delta_{\text{M}}\approx 0.2$ electron doping, which takes place when the Fermi level is lifted up to the flat top of the bonding $d_{z^2}$ band. Generally, larger DOS renders smaller $U_c$. However, the doping $\delta_{\text{m}}$ for the minimal $U_c$ is slightly lower than $\delta_{\text{M}}$, which takes place when the $\alpha_1$- and $\gamma$- pockets are best nested to each other. The $\delta$-dependence of $\lambda$ is shown in Fig.~\ref{dop}(b). With hole doping, the pairing symmetry maintains $s^{\pm}$ but the $\lambda$ and hence $T_c$ obviously drop; while with slight electron doping, the $\lambda$ and hence $T_c$ promptly arise. When the doping approaches near $\delta_{\text{m}}$, the system enters the SDW phase for $U=0.9$ eV.  

The distribution of the relative gap function on the FS for $\delta=0.1$ electron doping is shown in Fig.~\ref{dop}(c). The FS-nesting for this doping is much better than that for the undoped case shown in Fig.\ref{sc}(b), which is also reflected in the very sharp susceptibility peak at $\mathbf{Q}_1$ displayed in Fig.~\ref{dop}(d). The reason for the improved FS-nesting lies in that electron doping reduces the size of the $\gamma$- pocket while it enlarges the $\alpha_1$- pocket. Consequently, the $T_c$ for this doping is much higher than that for the undoped case. Mean while, the $s^{\pm}$-pattern of the pairing gap function maintains and is even more obvious. The real-space pairing pattern is still dominated by the interlayer pairing between the $d_{z^2}$ orbitals.

For $\delta=-0.1$ hole doping, the $\gamma$-pocket grows larger than that for the undoped case, while the $\alpha_1$ pocket vanishes,as shown in Fig.~\ref{dop}(e) (see  the solid lines for the FS). In this case, there is no nesting between the $\gamma$-pocket and the already vanished $\alpha_1$ pocket in the purely geometric sense. However, as the local bottom of the non-bonding $d_{z^2}$ band at the $\Gamma$-point is very close to the Fermi level, there still exists a spin-susceptibility peak locating at the vector $\mathbf{Q}_1$ connecting the $\Gamma$-point and the boundary of the $\gamma$-pocket, as shown in Fig.~\ref{dop}(f). In this sense, the vector $\mathbf{Q}_1$ can be viewed as a virtue ``FS-nesting'' vector which works in the pairing pattern. To show the effect of this virtual FS-nesting, we provide the distribution of the gap function within an energy shell of $\pm 0.01$eV around the Fermi level in Fig.~\ref{dop}(e). Consequently, the regime near the $\Gamma$-point still hosts the largest gap amplitude, although it is slightly away from the Fermi level. What's more, the gap sign in this regime is still opposite to that on the $\Gamma$-pocket. In this sense, we still get an $s^{\pm}$-wave pairing despite the vanish of the $\alpha_1$-pocket. We also note that for this doping level, two new FS-nesting vectors emerge: the $\mathbf{Q}_2$ between the $\gamma$- and $\beta_1$- pockets, and the $\mathbf{Q}_3$ between the $\alpha_2$- and $\beta_1$- pockets, which are reflected in the distribution of the spin susceptibility in Fig.~\ref{dop}(f) as different local maxima. Consequently, the nested patches between the $\gamma$- and $\beta_1$- pockets and those between the $\alpha_2$- and $\beta_1$ pockets also carry gap functions with opposite signs, but with weaker gap amplitude.

\begin{figure}[htbp]
\centering
\includegraphics[width=0.5\textwidth]{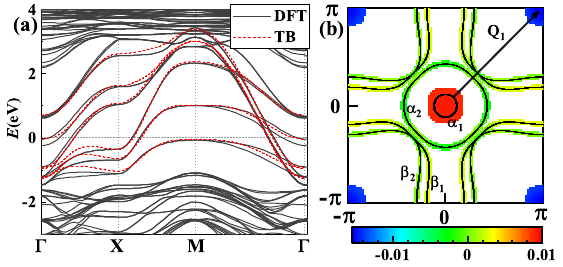}
\caption{(color online) (a) The DFT(black solid line) and TB(red dashed line) band structure of $\text{La}_4\text{Ni}_3\text{O}_{10}$ under the pressure of 40Gp, the atomic positions and crystal lattices were fully relaxed. (b) The distribution of the leading $s$-wave pairing gap function within an energy shell $\pm 0.03$ eV near the Fermi energy for $U=0.9$ eV. The FS pockets are marked by black lines. The ``virtual'' nesting vector $\mathbf{Q_1}$ connects the boundary of the $\alpha_1$ pocket and the $M$-point.}
\label{band2}
\end{figure}

{\bf Robust $s^{\pm}$-wave pairing:} As the top of the bonding $d_{z^2}$ band at the $M$-point and the local bottom of the non-bonding $d_{z^2}$ band at the $\Gamma$-point are very close to the Fermi level, a slight variation of the band structure may lead to appearance or disappearance of the $\alpha_1$ or the $\gamma$ pocket. In order to investigate the influence of these pockets on the superconducting pairing in the system, we built a new band structure with the lattice constants and atomic positions of this tetragonal $I4/mmm$ structure under 40 Gpa fully relaxed. The new band structure is shown Fig.~\ref{dft}(a), while the resultant FS is marked by the black solid lines in Fig.~\ref{band2}(b). Distinguished from the previous band structure in Fig.~\ref{dft}(a), this new band structure lacks the $\gamma$ pocket at the $M$-point, as the top of the bonding $d_{z^2}$ band at the $M$-point sinks slightly below the Fermi level. This band structure hosts the $\alpha_1$-pocket. The TB parameters for this band structure are provided in the SM.

Adopting the same multi-orbital Hubbard model as described in Eq.~\ref{model}, our RPA calculations  reveal that, despite vanishing of the $\gamma$-pocket, the maximum of the spin susceptibility still locates at the ``virtual'' nesting vector $\mathbf{Q}_1\approx(\pi,\pi)$ which connects the $M$-point and the boundary of the $\alpha_1$-pocket. This is due to two reasons: Firstly, the local band top at the $M$-point is very near the Fermi level; secondly, this band top is very flat leading to large DOS. Consequently, similarly with the case for $\delta=-0.1$ shown in Fig.~\ref{dop}(e), the strong spin fluctuation with momentum $\mathbf{Q}_1$ mediates the $s^\pm$-wave  pairing pattern shown in Fig.~\ref{band2}(b), in which the two regimes near the $\Gamma$- and $M$- points host strongest gap functions with opposite signs.  

Combining the results shown in Fig.~\ref{dop} (e) and Fig.~\ref{band2} (b), it becomes evident that in pressurized La$_4$Ni$_3$O$_{10}$, whether the $\gamma$-pocket or the $\alpha_1$-pocket vanishes or not, as long as they remain close to the Fermi level, the obtained gap function always takes the $s^{\pm}$ pattern, in which the two regimes near the $\Gamma$- and the $M$-points host strongest pairing gap with opposite signs.

{\bf Conclusion:} Adopting the TB model fitted from the DFT band structure, we have studied the pairing mechanism and pairing symmetry of pressurized La$_4$Ni$_3$O$_{10}$ by the RPA approach. Our results provide the $s^{\pm}$-wave pairing driven by spin fluctuations, with a $T_c$ obviously lower than that of pressurized La$_3$Ni$_2$O$_{7}$. The gap amplitude is dominantly distributed at the bottom regime of the non-bonding $d_{z^2}$ band and the top regime of the bonding $d_{z^2}$ band. These two regimes are connected by the nesting vector $\mathbf{Q}_1\approx (\pi,\pi)$, leading to the opposite signs between the gap functions within the two regimes. The real-space pairing pattern is dominated by the interlayer pairing between the $d_{z^2}$ orbitals.

\section*{Acknowledgements}
We are grateful to the discussions with Chen Lu. This work is supported by the NSFC under Grant Nos.12234016, 12074031, 12141402, and 12334002. W.-Q. Chen is supported by the National Key R\&D Program of China (Grants No. 2022YFA1403700), the Science, Technology and Innovation Commission of Shenzhen Municipality (No. ZDSYS20190902092905285), and Center for Computational Science and Engineering of Southern University of Science and Technology.


\end{document}